\documentclass[10pt]{iopart}
\usepackage{iopams,graphicx,color}  
\begin{document}

\title{Latest results from NA60} 

\def\cagl{$^{a}$}
\def\cern{$^{b}$}
\def\lpc{$^{c}$}
\def\heid{$^{d}$}
\def\lis{$^{e}$}
\def\llr{$^{f}$}
\def\riken{$^{g}$}
\def\suny{$^{h}$}
\def\torino{$^{i}$}
\def\ipnl{$^{j}$}
\def\yer{$^{k}$}

\author{A. F{\"o}rster for the NA60 Collaboration: \\  
R.~Arnaldi\torino, 
R.~Averbeck\suny, 
K.~Banicz\heid$^,$\cern, 
J.~Castor\lpc, 
B.~Chaurand\llr, 
C.~Cical\`o\cagl, 
A.~Colla\torino, 
P.~Cortese\torino, 
S.~Damjanovic\heid, 
A.~David\lis$^,$\cern, 
A.~De~Falco\cagl, 
A.~Devaux\lpc, 
A.~Drees\suny, 
L.~Ducroux\ipnl, 
H.~En'yo\riken,
J.~Fargeix\lpc,  
A.~Ferretti\torino, 
M.~Floris\cagl,
A.~F{\"o}rster\cern,  
P.~Force\lpc, 
N.~Guettet\cern$^,$\lpc, 
A.~Guichard\ipnl, 
H.~Gulkanian\yer, 
J.~M.~Heuser\riken, 
M.~Keil\lis$^,$\cern, 
L.~Kluberg\llr$^,$\cern,
C.~Louren\c{c}o\cern,
J.~Lozano\lis, 
F.~Manso\lpc, 
A.~Masoni\cagl, 
P.~Martins\lis$^,$\cern, 
A.~Neves\lis, 
H.~Ohnishi\riken, 
C.~Oppedisano\torino, 
P.~Parracho\cern, 
Ph.~Pillot\ipnl, 
G.~Puddu\cagl, 
E.~Radermacher\cern, 
P.~Ramalhete\cern, 
P.~Rosinsky\cern, 
E.~Scomparin\torino, 
J.~Seixas\lis$^,$\cern, 
S.~Serci\cagl, 
R.~Shahoyan\lis$^,$\cern, 
P.~Sonderegger\lis, 
H.~J.~Specht\heid$^,$\cern, 
R.~Tieulent\ipnl, 
G.~Usai\cagl, 
R.~Veenhof\lis$^,$\cern, 
H.~K.~W\"ohri\lis$^,$\cern
}

\address{
\cagl \mbox{Universit\`a di Cagliari and INFN, Cagliari, Italy}\\
\cern \mbox{CERN, Geneva, Switzerland}\\
\lpc \mbox{LPC, Universit\'e Blaise Pascal and CNRS-IN2P3, Clermont-Ferrand, France}\\
\heid \mbox{Universit\"{a}t Heidelberg, Heidelberg, Germany}\\
\lis \mbox{CFTP, Instituto Superior T\'ecnico, Lisbon, Portugal}\\
\llr \mbox{LLR, Ecole Polytechnique and CNRS-IN2P3, Palaiseau, France} \\
\riken \mbox{RIKEN, Wako, Saitama, Japan}\\
\suny \mbox{SUNY, Stony Brook, NY, USA}\\
\torino \mbox{Universit\`a di Torino and INFN, Turin, Italy}\\
\ipnl \mbox{IPNL, Universit\'e Claude Bernard Lyon-I and CNRS-IN2P3, Villeurbanne, France}\\
\yer \mbox{YerPhI, Yerevan, Armenia}\\
}

\ead{andreas.foerster@cern.ch}

\begin{abstract}
The NA60 experiment has measured the production of  muon pairs
and of charged particles in \mbox{In+In} collisions 
at a beam energy of $158$~$A$~GeV.
For invariant dimuon masses below the $\phi$ 
the space-time averaged $\rho$ spectral function 
was isolated by a novel procedure. 
It shows a strong broadening but essentially no shift in mass.
The production of 
\mbox{J/$\psi$} was measured as a function of the collision
centrality. As in previous experiments 
studying \mbox{Pb+Pb} collisions an anomalous
supression is observed, setting in at approximately
90 participant nucleons.
Using the charged particles 
the reaction plane was reconstructed.
The elliptic flow of charged particles 
increases with
$p_{\rm t}$ showing a saturation for $p_{\rm t} > 2$~GeV/$c$. 
For the first time azimuthal distributions for \mbox{J/$\psi$}
are shown.
\end{abstract}


\maketitle


\section{Introduction} \label{intro}

High-energy heavy-ion collisions provide a unique
opportunity to study the behaviour of strongly 
interacting matter at high densities and at high 
energy densities. They are the only way to investigate 
a possible phase transition from hadronic matter
to a plasma of deconfined quarks and gluons
as well as  the restoration of the chiral symmetry
which is spontaneously broken in the hadronic world.
These issues have been studied by several experiments
at the SPS (Super Proton Synchrotron) at CERN
and many exciting results were obtained.
NA60 is a second generation 
experiment designed to answer specific questions
that still remain open after the completion of these 
previous experiments.

The main focus of NA60 is the study
of muon pair production.
The experimental setup is based on the muon spectrometer previously used
by NA38 and NA50~\cite{na50setup}, separated from the target region
by a $5.5$~m long hadron absorber (mostly carbon)
and composed of a toroidal magnet, 
of eight multi-wire proportional chambers
and of four scintillator trigger telescopes. 
A Zero-Degree Calorimeter (ZDC) measures the energy
of the spectator nucleons to determine the
collision centrality.
The muon spectrometer is complemented by a 
high-granularity silicon pixel tracker of
unprecedented radiation tolerance~\cite{Gluca:2005,Keil:2005zq}
which was constructed using  ALICE sensors 
and readout chips~\cite{Riedler:2006}. 
The tracker, embedded in a 2.5~T dipole magnet in the vertex region,
tracks all charged particles before the hadron absorber and determines
their momenta independently of the muon spectrometer. 
It allows for a determination of the primary interaction vertex
with an accuracy of $~10-15$~$\mu$m in the transverse plane
and $~200$~$\mu$m along the beam axis.
The matching of the muon tracks before and after the 
hadron absorber, both in angular and in momentum space, 
improves the dimuon mass resolution in
the region of the vector mesons $\omega$ and $\phi$ to
$\sim$20~MeV/c$^{2}$, significantly reduces the combinatorial background due
to $\pi$ and K decays and makes it possible to measure the muon
offset with respect to the interaction vertex~\cite{Ruben:2005qm}. 
Measuring all charged particles within the vertex tracker acceptance
allows in addition for the reconstruction of the direction
of the impact parameter vector and hence to determine 
the orientation of the reaction plane and to study 
azimuthal angle distributions of the emitted particles.

The experiment has taken data in 2003 (\mbox{In+In} collisions)
and in 2004 (proton+nucleus collisions).
The results reported upon in this paper were obtained from the analysis of
the data taken in 2003 with a 158 AGeV Indium beam, incident on a
segmented Indium target of seven disks with a total of 18\% 
nuclear interaction length. At an average beam intensity 
of 5$\cdot$10$^{7}$
ions per 5~s burst, about $3\cdot10^{12}$ ions were delivered to the
experiment and a total of $230\cdot10^6$ dimuon triggers were recorded.


\section{The $\boldsymbol{\rho}$ spectral function} 
\label{rho}

In the low-mass sector ($m < m_{\phi}$) the CERES experiment 
has studied the production of electron pairs in 
\mbox{p+Be/Au}, \mbox{S+Au} and \mbox{Pb+Au}
collisions~\cite{Aga05}. 
In nuclear collisions a clear excess above the expected hadronic
sources has been observed in the dielectron mass distribution. 
The origin of the excess has been commonly 
interpreted in terms of thermal production from the dense hadronic gas created
in the collision, mainly occurring via the $\pi^+ \pi^- \rightarrow \rho
\rightarrow {\rm e^+} {\rm e^-}$ process. Even if it is clear that an in-medium
modification of the $\rho$ must be introduced in 
order to explain the results,
the lack of statistics and mass resolution have prevented any detailed
understanding of the character of the in-medium changes.

\begin{figure}[h]
\begin{center}
\includegraphics*[width=6cm,clip=, bb = 6 12 560 665]{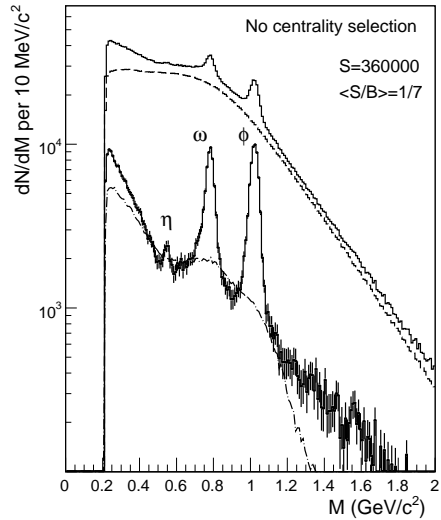}
\vspace*{-0.2cm}
\caption{Mass distribution of opposite-sign dimuons (upper
  histogram), combinatorial background (dashed), signal fake
  matches (dashed-dotted), and resulting signal 
  (lower histogram with error bars).}
\label{fig_lmr_massspec}
\end{center}
\vspace*{-0.4cm}
\end{figure}

Figure~\ref{fig_lmr_massspec} shows the 
opposite-sign dimuon
mass distribution as measured by NA60, integrated over 
all collision centralities. 
The combinatorial background of uncorrelated muon pairs mainly
originating from $\pi$ and K decays is determined using a
mixed-event technique~\cite{Ruben:2005qm}.
After subtraction the remaining opposite-sign muon pairs 
still contain ``signal'' fake matches (associations of muons
measured in the muon spectrometer to non-muon tracks in the vertex tracker), 
a contribution which is only 7\% of the combinatorial
background level. It has been determined by an
overlay Monte Carlo method. 
After subtracting the combinatorial background and the signal fake matches,
the resulting net spectrum contains about 360\,000 muon pairs in the
mass range of the figure, roughly 50\% of the total available
statistics.  
The analysis is done in
four classes of collision centrality defined by the
charged particle multiplicity density: 
peripheral (4-30),
semiperipheral (30-110), semicentral (110-170) and central (170-240).
Numerically, the average rapidity density in  each 
class is, within $10\%$, equal
to the average number of participants in  the class.
The signal-to-background ratios associated with these classes
are 2, 1/3, 1/8 and 1/11, respectively.

The most peripheral data can be described
on the basis of known sources, i.e.\ the two-muon  
and the Dalitz decays of the various vector mesons,
plus a contribution from open charm decays. 
The expected mass 
shape for the various physics processes has been obtained propagating decay
muons through the NA60 set-up, using GENESIS~\cite{genesis} 
(originally based on~\cite{genesis_ceres})
as the event generator and GEANT for tracking.
This `hadronic' cocktail has been directly fitted to the data and 
is found to reproduce the observed spectrum 
quite well~\cite{Damjanovic:qm2005}.

\begin{figure}[t!]
\centering
\begin{minipage}[t]{0.48\textwidth}
\includegraphics*[width=6cm,clip=]{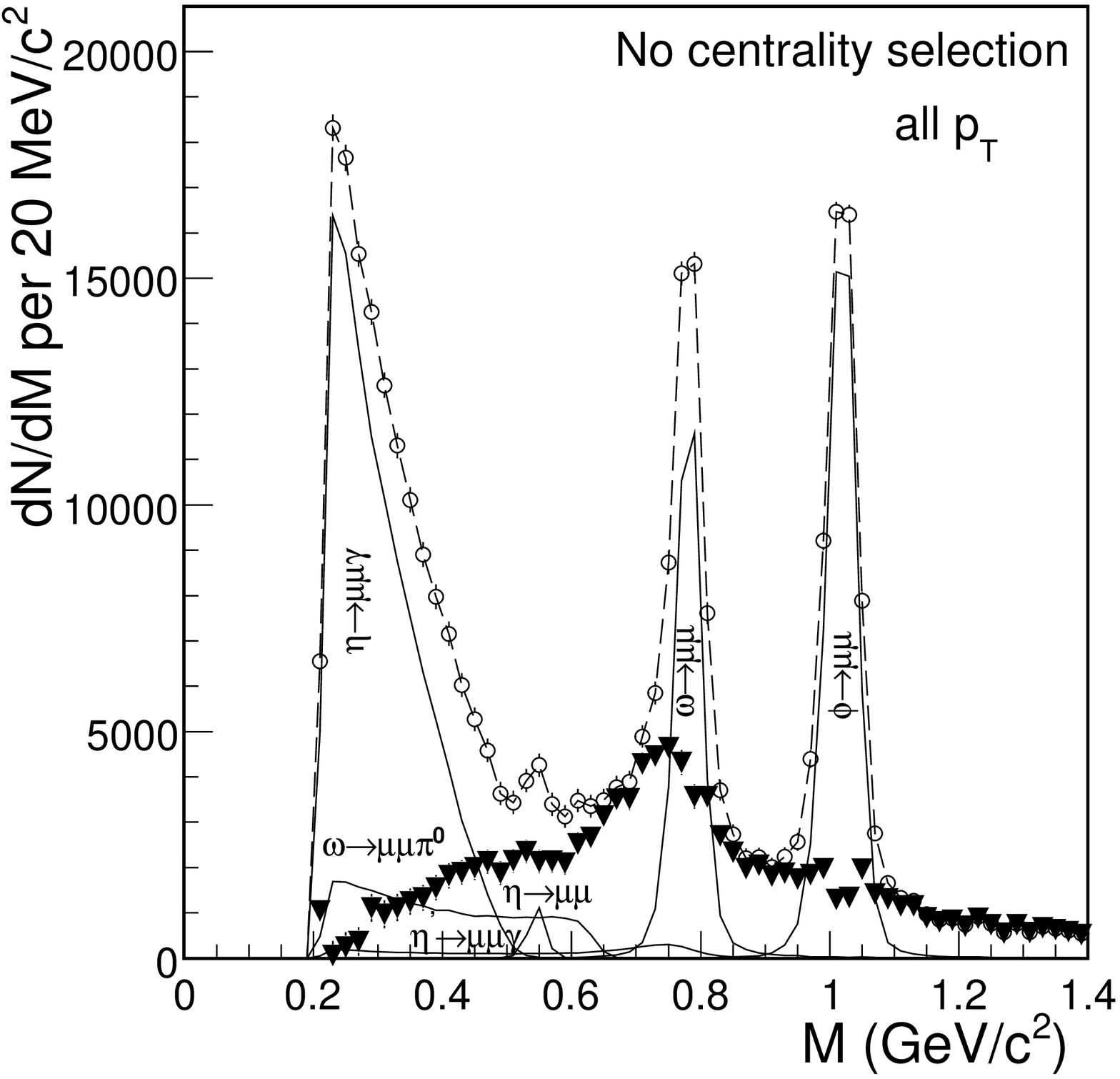}
\vspace*{-0.2cm}
\caption{Isolation of an excess above the electromagnetic decays
  of a ``conservative'' hadron decay cocktail (see text). Total
  data (open circles), individual cocktail sources (solid), difference
  data (thick triangles), sum of cocktail sources and difference data
  (dashed).}
\label{fig_lmr_subtr_sources}
\end{minipage}
\begin{minipage}[t]{0.48\textwidth}
\includegraphics*[width=6cm,clip=]{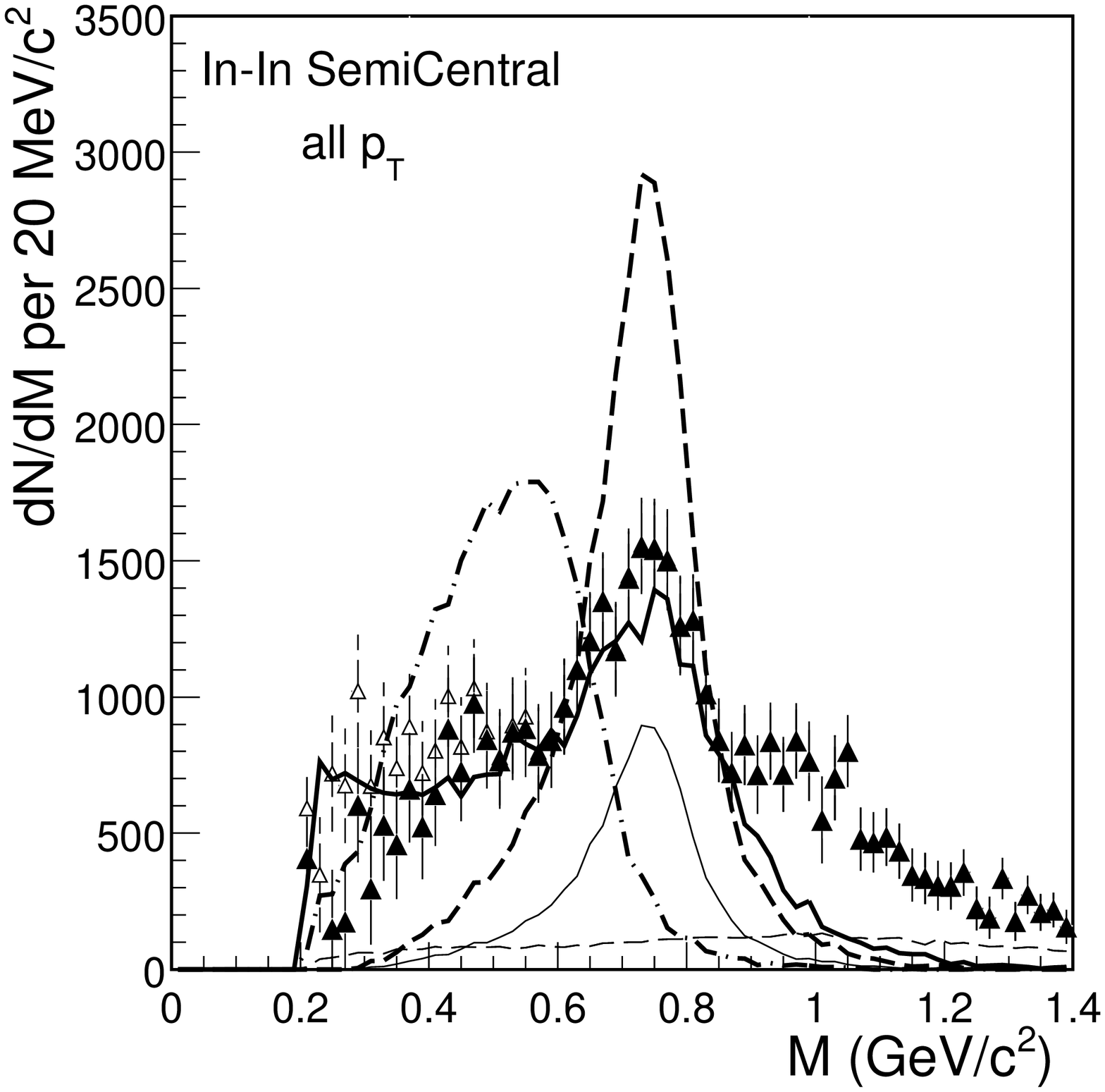}
\vspace*{-0.2cm}
\caption{Comparison of the excess mass spectrum for the semi-central
  bin to model predictions, made for In-In at
  ${\rm d}N_{\rm ch}/{\rm d}\eta$=140. 
  Cocktail $\rho$ (thin solid), unmodified $\rho$
  (dashed), in-medium broadening $\rho$~\cite{Rapp:1995zy,Rapp:1999ej}
  (thick solid), in-medium moving $\rho$ 
  related to~\cite{Brown:kk,Brown:2001nh}
  (dashed-dotted). 
}
\label{fig_lmr_excess}
\end{minipage}
\vspace*{-0.4cm}
\end{figure}

In the more central bins, the data can no longer be described on the
basis of the standard hadron decay cocktail alone, but are indicative
of the existence of an excess yield. Since the particle ratios are
expected to be different from the peripheral data, global fits to the
more central data are bound to bias both the extracted
cocktail parameters and an excess with a priori unknown
characteristics. We have therefore used a novel procedure
which is illustrated in
Fig.~\ref{fig_lmr_subtr_sources} and described in detail 
in~\cite{prl_lmr}. 
The excess is isolated by subtracting the
cocktail, without the $\rho$, from the data. The cocktail is fixed,
separately for the major sources and in each centrality bin, by a
``conservative'' approach. The yields of the narrow vector mesons
$\omega$ and $\phi$ are fixed so as to get, after subtraction, a
smooth underlying continuum. For the $\eta$, an upper limit is
defined by ``saturating'' the measured data in the region close to 
$0.2$~GeV/$c^{2}$; this implies the excess to vanish at very low mass by
construction.  

The excess mass spectrum resulting from the
subtraction of the ``conservative'' hadron decay cocktail from the
measured data is shown in Fig.~\ref{fig_lmr_excess} for semi-central
collisions. The
qualitative feature of the spectrum is striking: a peaked structure
is seen  around the position of the nominal $\rho$ pole.
The excess spectrum is 
consistent with an interpretation of the excess as being dominated by
$\pi\pi$ annihilation. 

Figure~\ref{fig_lmr_excess} as well shows predictions from theoretical 
calculations employing the  broadening scenario
of~\cite{Rapp:1995zy,Rapp:1999ej} and the moving-mass scenario related
to~\cite{Brown:kk,Brown:2001nh}. Both are evaluated for \mbox{In+In} at
${\rm d}N_{\rm ch}/{\rm d}\eta = 140$ 
within the same fireball evolution.
In addition the unmodified $\rho$ is shown.
The data shown in this paper have not been corrected for the mass-
and $p_{\rm t}$-dependent acceptance of the NA60 setup. The theoretical
calculations were therefore propagated
through the acceptance filter to allow for a fair comparison with the
data. The integrals of the theoretical spectra are 
normalized to the data in
the mass interval $M < 0.9$~GeV/$c^{2}$. The unmodified $\rho$ is
clearly ruled out. The specific moving-mass scenario plotted here,
which fitted the 
CERES data~\cite{Aga05,Rapp:1999ej,Agakichiev:1997au}, is
ruled out as well.
The broadening scenario appears more realistic. However,
the nearly symmetrical broadening around the $\rho$ pole seen in the
data is not reproduced by this model. The remaining excess at 
$M>0.9$~GeV/$c^{2}$ may well be related to the prompt dimuon excess found by
NA60 in the intermediate mass region~\cite{Ruben:2005qm}. Processes
other than \mbox{2$\pi$}, i.e. \mbox{4$\pi$ ...} 
could possibly account for
the region $M > 0.9$~GeV/$c^{2}$~\cite{gale:nn}.


\section{The anomalous \mbox{J/$\boldsymbol{\psi}$} supression} 
\label{jpsi}

In \mbox{Pb+Pb} collisions above a certain centrality threshold an anomalous 
\mbox{J/$\psi$} suppression has been seen by NA50~\cite{na50jpsi}, 
i.e.\ suppression mechanisms
different from nuclear absorption must be invoked to explain 
the observed \mbox{J/$\psi$} yield. However, several questions raised by 
this observation have still to be clarified.
By studying the \mbox{J/$\psi$} production in 
\mbox{In+In} collisions NA60 
investigates the onset of the anomalous \mbox{J/$\psi$} behaviour in systems
lighter than \mbox{Pb+Pb}.
By comparing the suppression pattern obtained in
different systems as a function of various centrality variables it should
be possible to single out a scaling variable for the anomalous 
\mbox{J/$\psi$} suppression.
\begin{figure}[h]
\begin{minipage}[t]{0.48\textwidth}
\centering 
\includegraphics[width=6.0cm]{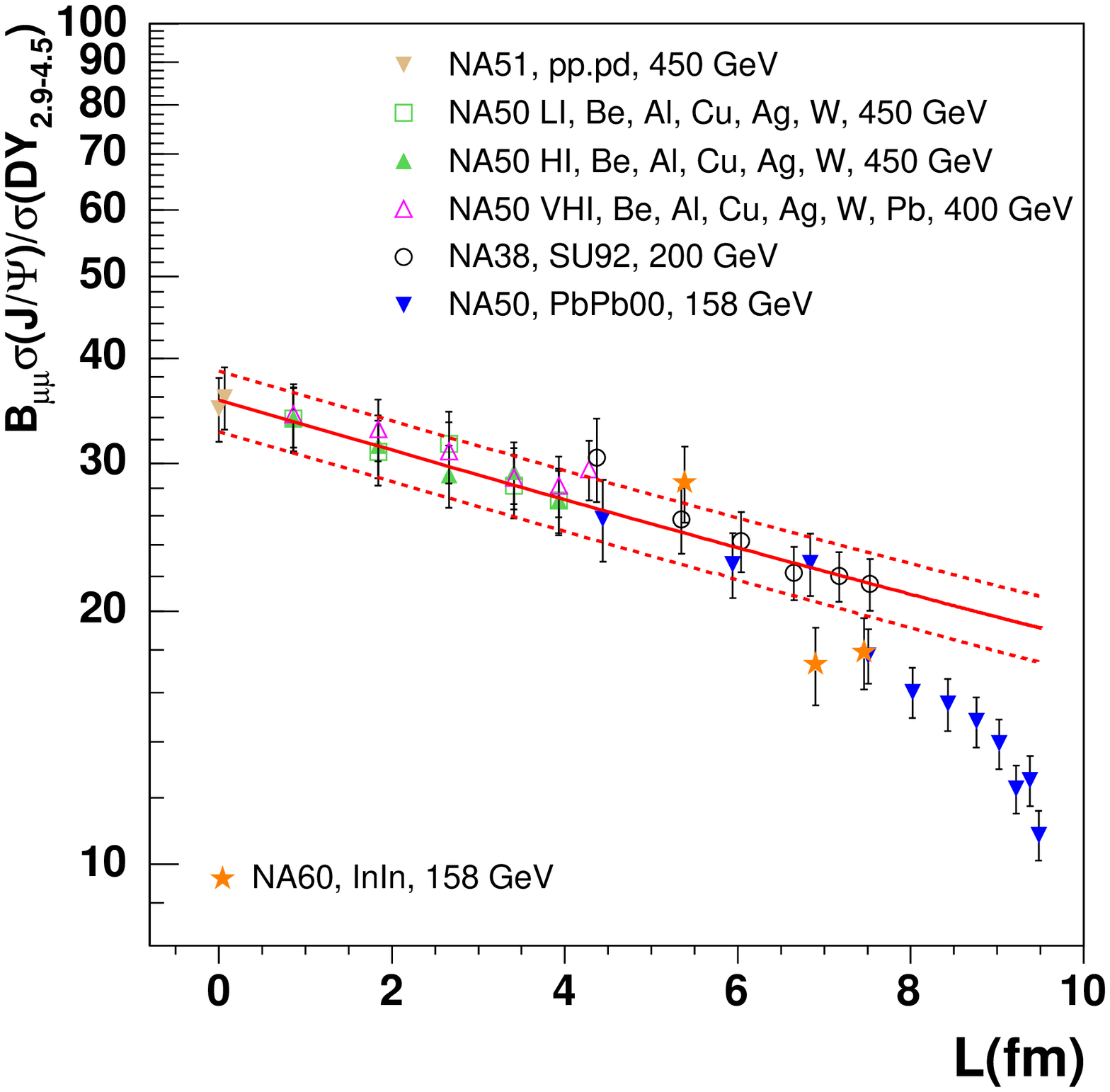}
\vspace*{-0.2cm}
\caption{The \mbox{J/$\psi$} over DY standard analysis as a function of $L$ (see text), compared to the results of 
NA50/NA38~\cite{na50jpsi}.}
\label{fig_jpsi_l}
\end{minipage}
\begin{minipage}[t]{0.48\textwidth}
\centering 
\includegraphics[width=5.9cm]{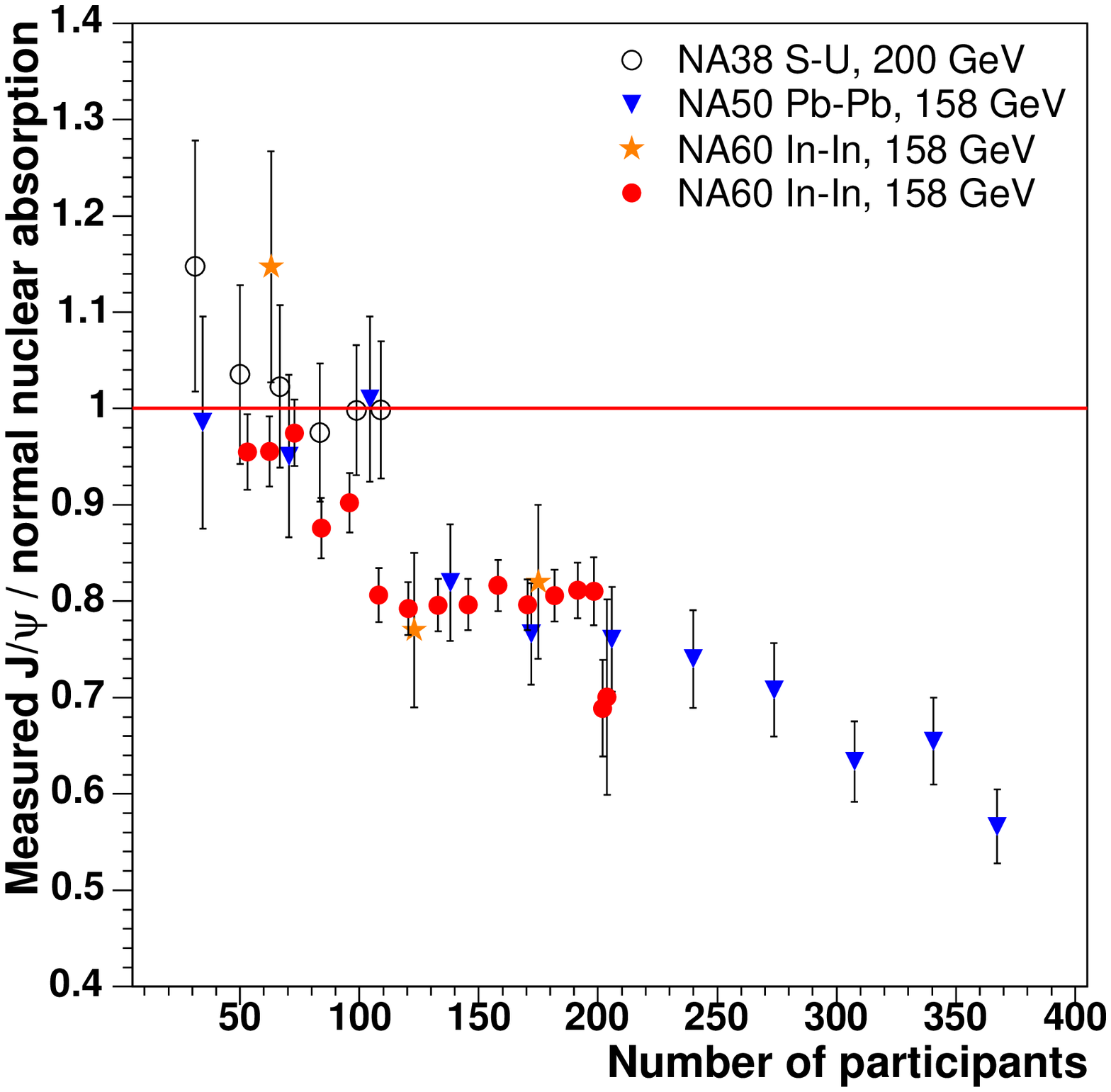}
\vspace*{-0.2cm}
\caption{\mbox{J/$\psi$} suppression pattern measured in
\mbox{S+U}, \mbox{In+In} and \mbox{Pb+Pb}, as a function of $N_{\rm part}$.}
\label{fig_jpsi_npart}
\end{minipage}
\vspace*{-0.2cm}
\end{figure}
Two different analyses have been performed by NA60 
to investigate the 
centrality dependence of the \mbox{J/$\psi$} production,
corresponding to two different ways of normalizing its yield~\cite{ArnQM}. 
The first technique, the so-called `standard analysis', 
is based on a normalization to 
the Drell-Yan events, already widely used in the past by NA38/NA50. 
The study of the ratio between the \mbox{J/$\psi$} and the 
Drell-Yan cross sections has the
advantage of being free from systematic errors connected with the efficiency
and luminosity evaluations. 
However, the statistical error is large due to the
small number of high-mass Drell-Yan pairs (a few hundreds); 
therefore only three centrality bins 
can be defined using the energy measured  
in the ZDC ($E_{\rm ZDC}$).
Figure~\ref{fig_jpsi_l} shows the result as a function of the 
length $L$ of nuclear matter crossed by the charmonium state
in comparison to the results of NA38/NA50.

The second technique overcomes the statistical problem 
by directly studying 
the measured \mbox{J/$\psi$} centrality distribution
as a function of $E_{\rm ZDC}$ and correcting the 
inefficiencies due 
to the reconstruction procedure and the event selection
which only affect very peripheral events
($E_{\rm ZDC} > 15$~TeV) and which are very small ($<1\%$). 
This distribution is
compared to the expected pattern if nuclear
absorption was the only active suppression mechanism. 
This reference curve is
obtained with the Glauber model, assuming that \mbox{J/$\psi$}
production is a hard process, and using $\sigma_{\rm abs} = 4.18$~mb for the
nuclear absorption cross section of the 
\mbox{$\rm c\overline c$} pair~\cite{Bor05}.
In such an analysis, the relative normalization between data and the reference 
curve is not fixed a priori. We have therefore imposed the 
centrality-averaged ratio between data and reference to be equal to the one
obtained in the standard analysis. Since the average 
$\sigma_{\rm J/\psi}/\sigma_{\rm DY}$ ratio has been 
measured with a 7\% statistical
error, such an overall error (not shown in the figure) applies to 
the results. 

In Fig.~\ref{fig_jpsi_npart} the \mbox{J/$\psi$} 
suppression pattern measured in
\mbox{In-In} collisions is plotted as a function of 
the number of participants $N_{\rm part}$ as determined
from $E_{\rm ZDC}$
(see~\cite{npart_calc} for details). 
We take into account the experimental resolution of the detector
as well as the influence of fluctuations in $N_{\rm part}$
at fixed impact parameter.
The suppression pattern 
is compared to the results obtained 
for \mbox{S+U} and
for \mbox{Pb+Pb} by 
NA50/NA38~\cite{na50jpsi}.
The observed pattern indicates that a suppression
is already present in \mbox{In+In} collisions, setting in at 
$\sim 90$ participating nucleons.
Even if $N_{\rm part}$ seems to be a reasonable scaling variable 
for the anomalous
suppression it is fair to say that more accurate data for the
previously studied collision
systems are needed in order to establish a firm conclusion on 
a scaling variable for the suppression mechanism.
For comparisons as a function of other centrality related variables 
and for comparisons to theoretical predictions see~\cite{ArnQM}.

\section{Elliptic flow of charged hadrons and 
of \mbox{J/$\boldsymbol{\psi}$}} 
\label{flow}

\begin{figure}[h]
\begin{center}
\vspace*{-0.4cm}
\includegraphics[width=6cm,clip=]{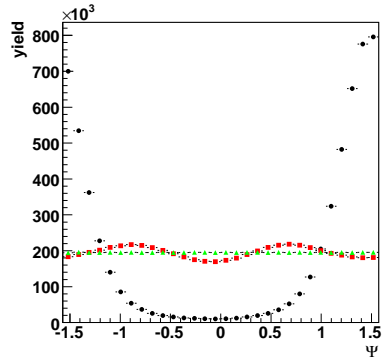}
\vspace*{-0.4cm}
\caption
{Distribution of the reconstructed event plane angles.
Black circles: without any correction, red squares: after recentering,
green triangles: after flattening.}
\label{fig_event_plane}
\vspace*{-0.5cm}
\end{center}
\end{figure}

In non-central heavy-ion collisions collective flow leads to characteristic
azimuthal correlations between particle momenta and the reaction
plane. This is the plane defined by the beam direction $\vec{z}$
and the impact parameter vector $\vec{b}$.
To quantify the anisotropic flow the coefficients
of a Fourier expansion of the azimuthal distributions with respect 
to the reaction plane are evaluated.
The reaction plane is experimentally not directly
accessible but it can be determined using the anisotropic flow 
itself, independently for each harmonic $n$ of the anisotropic
flow. A detailed overview over several methods is 
given in~\cite{poskanzer}. 
The acceptance coverage of NA60 is mainly close to
midrapidity ($y \approx 3-4$).
Since the directed 
flow, meaning the first harmonic ($n=1$), 
is zero at midrapidity due to symmetry reasons we concentrate 
on the second harmonic ($n=2$) only. 
The method we employ is the so-called 
event plane method~\cite{poskanzer}, relating the 
azimuthal emission angles $\phi_{\rm i}$ of all the
charged particles $i$  
to the azimuthal angle ${\it\Psi}$
of the event plane (an estimate of the reaction plane) 
by the so-called event flow vector $\vec{Q}$
\begin{equation}
\label{eq_qvec}
\hspace{-1.5cm}
Q_{\rm x} =  Q \cos(2 {\it\Psi})  =  
\sum_{\rm i} p_{\rm t}^{\rm i} \cos(2 \phi_{\rm i}) \quad , \quad
Q_{\rm y} =  Q \sin(2 {\it\Psi}) = 
\sum_{\rm i} p_{\rm t}^{\rm i} \sin(2 \phi_{\rm i}) 
\end{equation} 
which allows to calculate 
\begin{equation}
\label{eq_psi}
{\it\Psi} = 
\frac{1}{2} \cdot \left( \tan^{-1} 
\frac{Q_{\rm y}}{Q_{\rm x}} \right)  =
\frac{1}{2} \cdot \left( \tan^{-1} 
\frac{\sum_{\rm i} p^{\rm i}_{\rm t} \sin(2 \cdot \phi_{\rm i}) }
{\sum_{\rm i} p^{\rm i}_{\rm t} \cos(2 \cdot \phi_{\rm i}) } \right) \quad .
\end{equation}  
The NA60 experiment has an acceptance which is highly asymmetric
in the azimuthal angle $\phi$. This strongly 
affects the distribution of reconstructed event plane angles ${\it\Psi}$
as demonstrated by the black circles in Fig.~\ref{fig_event_plane}.
For perfectly symmetric acceptances and efficiencies this distributions
has to be flat for the average over many events.
To correct for this asymmetries the event flow vector
$\vec{Q}_{n}$ is recentered in a 
$(p_{t},y)$-acceptance matrix by its mean value for each bin
of this matrix as averaged over many events \cite{na49}.
This changes Eq.~\ref{eq_psi} to:
\begin{equation}
\label{eq_psi_rec}
{\it\Psi} = 
\frac{1}{2} \cdot \left( \tan^{-1} 
\frac{\sum_{\rm i} p^{\rm i}_{\rm t} \left[ 
\sin(2 \cdot \phi_{\rm i}) - 
\langle \, \sin(2 \cdot \phi) \, \rangle_{\rm p_{t},y} \, \right] }
{\sum_{\rm i} p^{\rm i}_{\rm t} \left[ 
\cos(2 \cdot \phi_{\rm i})  - 
\langle \, \cos(2 \cdot \phi) \, \rangle_{\rm p_{t},y} \, \right] } 
\right) \quad .
\end{equation}  
The resulting event plane distribution 
is shown by the red squares in Fig.~\ref{fig_event_plane}.
By definition the recentering can only remove anisotropies
in the same harmonic as the one used for the reconstruction 
of the event plane. To remove remaining anisotropies in 
higher orders a flattening procedure has been applied \cite{poskanzer}.
This leads to a flat event plane distribution
as shown by the green triangles in Fig.~\ref{fig_event_plane}.

\begin{figure}[h]
\vspace*{-0.4cm}
\begin{minipage}[t]{0.48\textwidth}
\centering 
\includegraphics[width=6cm,clip=]{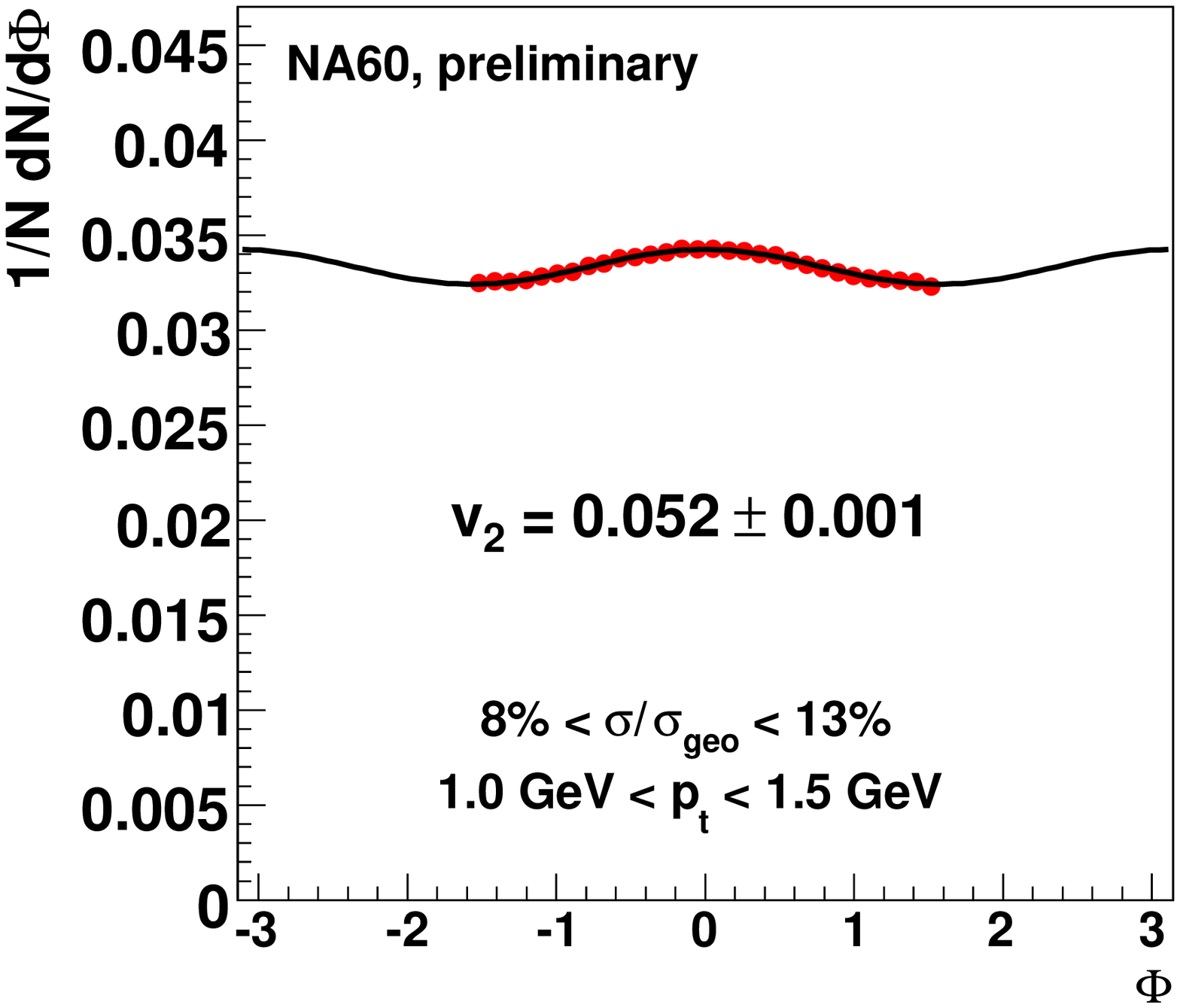}
\vspace*{-0.4cm}
\caption
{Azimuthal emission pattern of charged particles in \mbox{In+In}
at $158$~$A$~GeV. The line represents a fit to the data
according to Eq.~\ref{eq_fit_v2}.}
\label{fig_phi_fit_MB}
\end{minipage}
\begin{minipage}[t]{0.48\textwidth}
\centering 
\includegraphics[width=6cm,clip=]{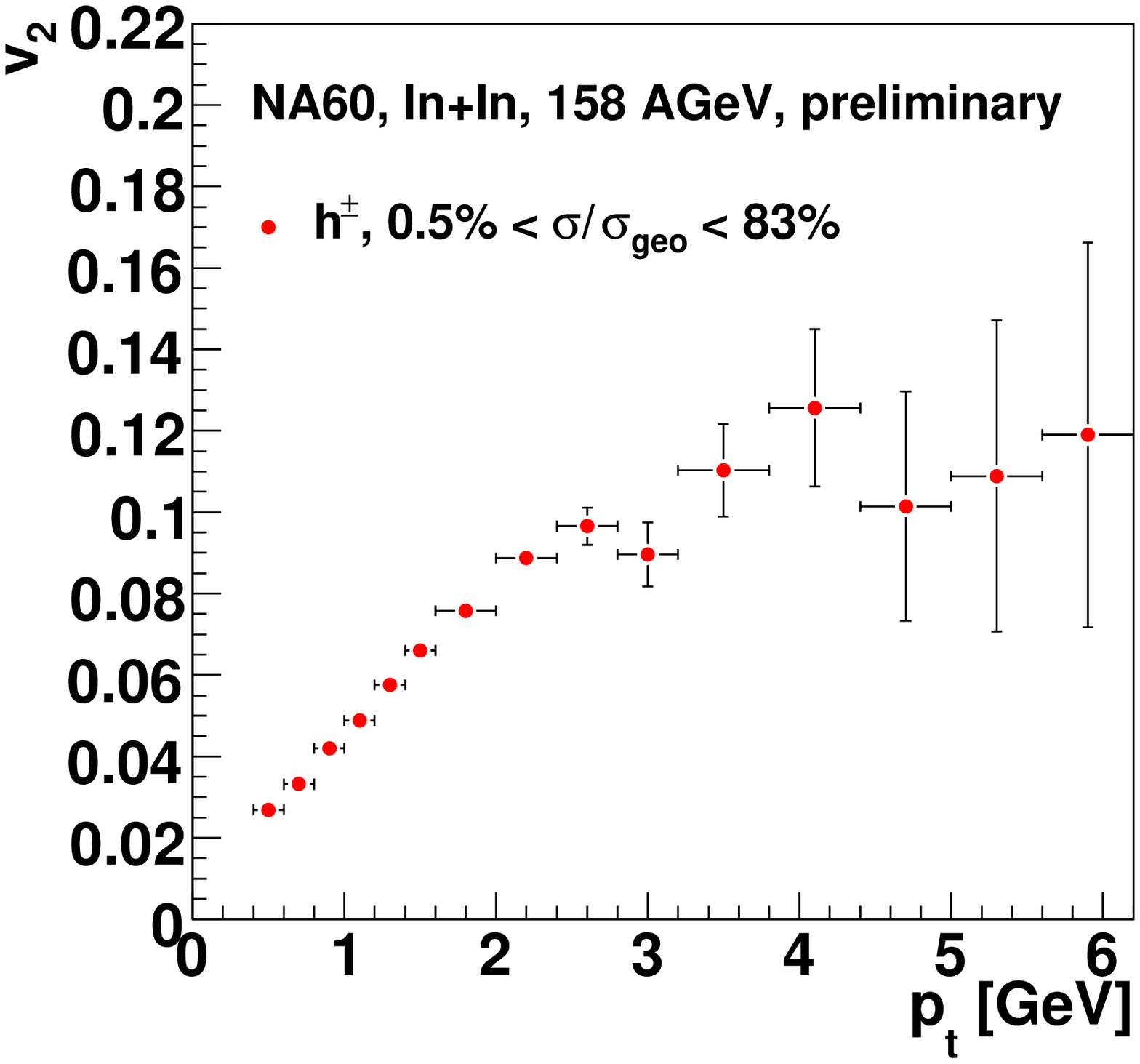}
\vspace*{-0.4cm}
\caption{The elliptic flow coefficient $v_2$
of charged hadrons as a function of $p_{\rm t}$.}
\label{fig_v2_pt_varbin_all}
\end{minipage}
\vspace*{-0.2cm}
\end{figure}
Once the bias in the event plane distribution
is removed it is possible to correlate
the emission angles of single particles
with this event plane and to create azimuthal
emission patterns ${\rm d}N/{\rm d}\phi$, with 
$\phi \, = \phi_{\rm lab} - {\it\Psi}$ being the difference
between the emission angle  $\phi_{\rm lab}$ of the particle
as measured in the laboratory and the orientation
of the event plane $\Psi$. To avoid autocorrelations
the event plane has to be recalculated 
without using the particle track whose emission
angle with respect to the event plane is being determined.
Figure~\ref{fig_phi_fit_MB} shows such an azimuthal emission pattern
for charged particles.
The line denotes the function
\begin{equation}
\label{eq_fit_v2}
\frac{{\rm d}N}{{\rm d}\phi} \, \sim \, 1 + \left[
2 \cdot v'_{2} \cdot \cos (2 \cdot \phi) \right]
\end{equation}
which was fitted to the data with 
$v'_{2} \, = \, \langle \cos (2 \cdot \phi) \rangle$.
This coefficient has to be corrected for 
the event plane resolution to obtain 
the real elliptic flow coefficient
$v_2 \, = \, 
v'_{2} / \langle \cos \left[2 \cdot {\it \Delta\Psi} \right] \rangle$.
%
This resolution has been determined by dividing each
event in two independend subevents and applying the
method described in~\cite{ollitrault}.
Depending on the collision centrality the event plane resolution
varies between $ 0.18 < \langle \cos \left[2 \cdot 
{\it \Delta\Psi} \right] \rangle < 0.32 $.

\begin{figure}[h]
\vspace*{-0.4cm}
\begin{minipage}[t]{0.48\textwidth}
\centering 
\includegraphics[width=6cm,clip=]{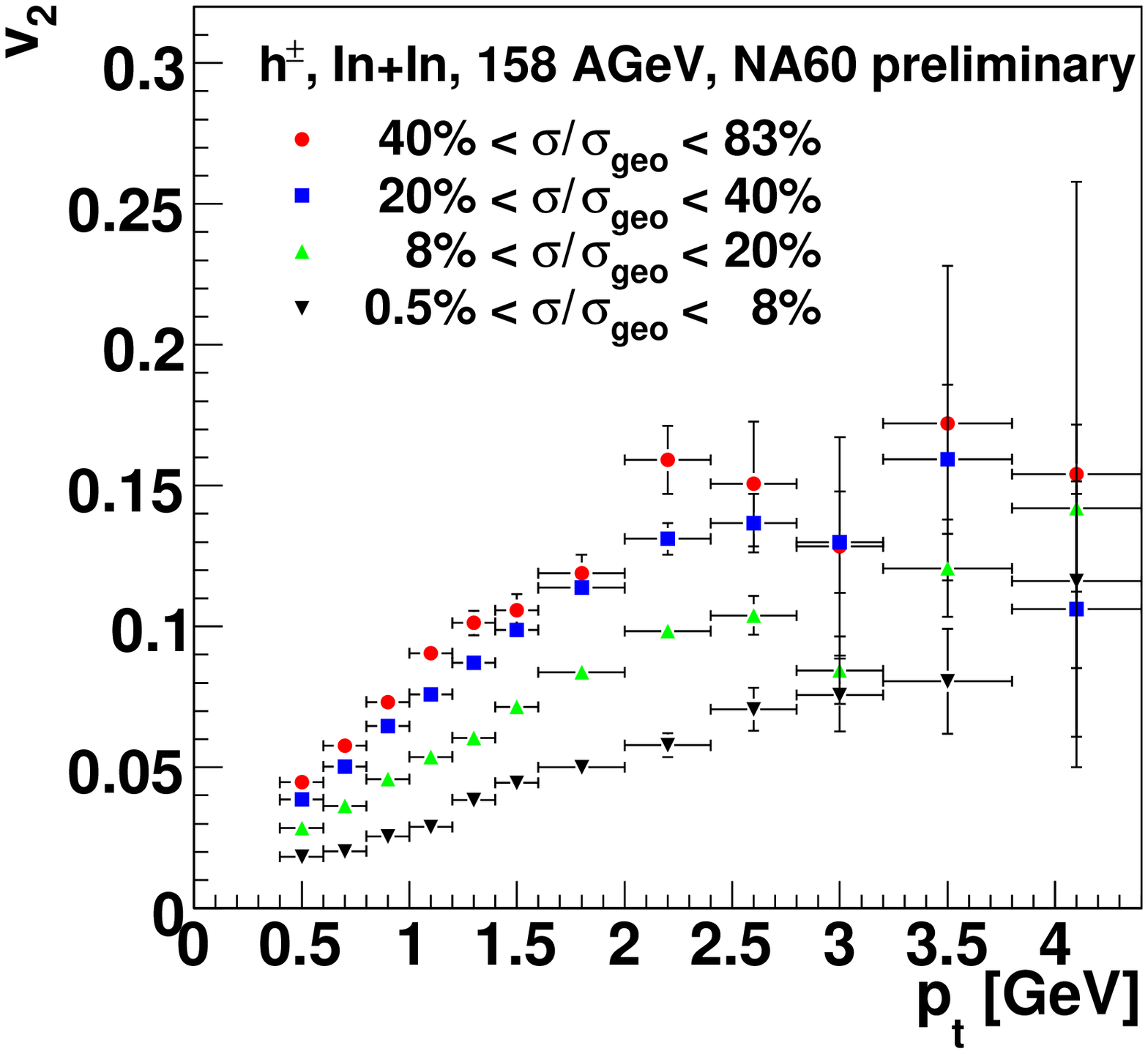}
\vspace*{-0.4cm}
\caption{The elliptic flow coefficient $v_2$
of charged hadrons as a function of $p_{\rm t}$ for 
various collision centralities.}
\label{fig_v2_pt_varbin_cent}
\end{minipage}
\begin{minipage}[t]{0.48\textwidth}
\centering 
\includegraphics[width=6cm,clip=]{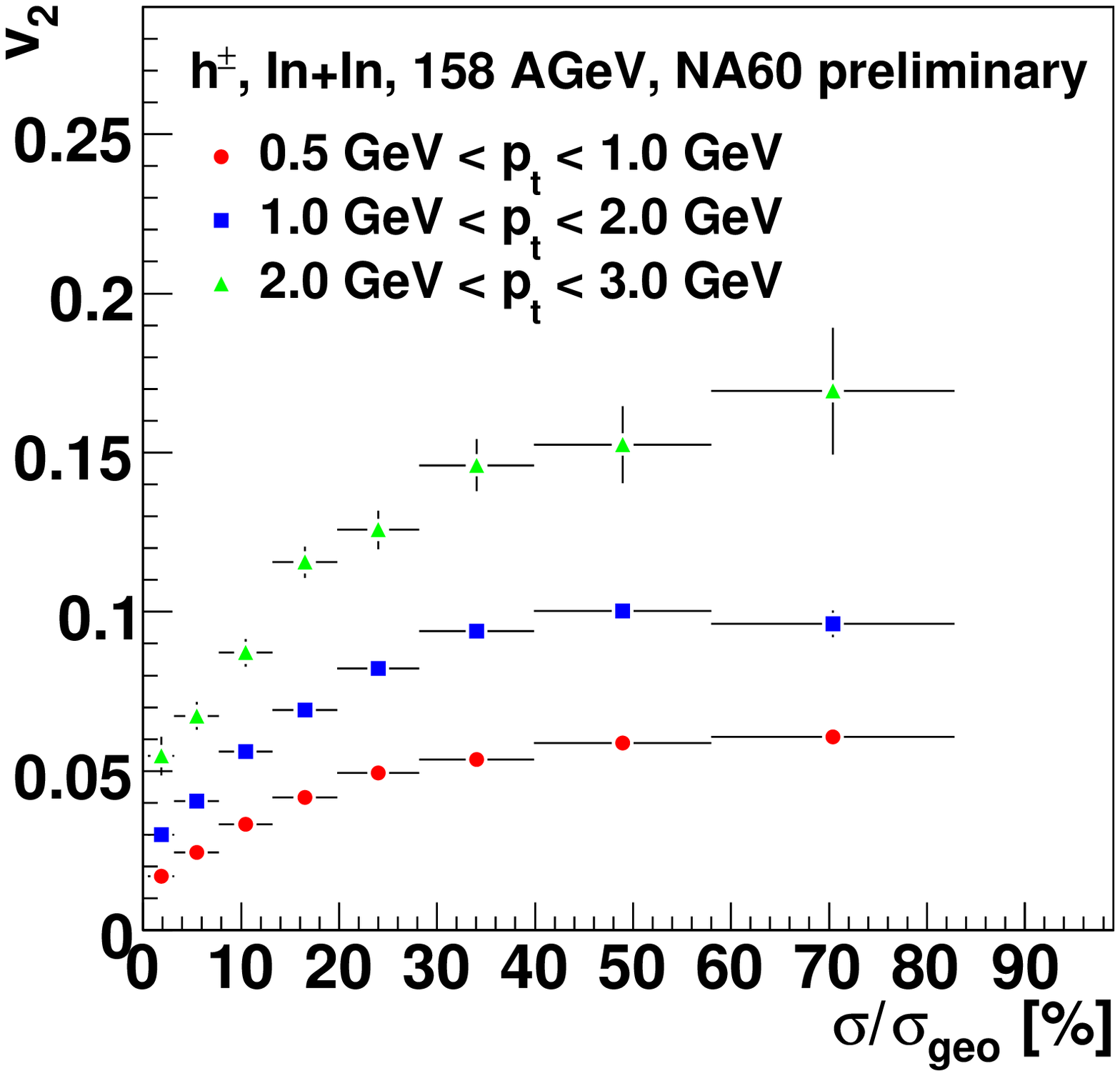}
\vspace*{-0.4cm}
\caption{The elliptic flow coefficient $v_2$
of charged hadrons as a function of the collision centrality  
for different ranges in $p_{\rm t}$.}
\label{fig_v2_cent_varbin}
\end{minipage}
\vspace*{-0.2cm}
\end{figure}
The resulting elliptic flow coefficients $v_2$ for 
all charged particles 
are displayed as a function of $p_{\rm t}$ 
in Fig.~\ref{fig_v2_pt_varbin_all}.
Please note that the present results are not corrected
for non-flow correlations due to the HBT effect.
Since this mainly affects the results at low $p_{\rm t}$
(see e.g.~\cite{dinh00}), we only present results
for $p_{\rm t} > 0.5$~\mbox{GeV/$c$}. 
Up to now all results shown in this section
are based on an analysis of approximately 50\% of
the measured statistics.
The flow coefficients rise with $p_{\rm t}$ and 
show a saturation at higher momenta as already 
previously observed in \mbox{Pb+Pb} collisions
at the SPS~\cite{v2_sps} 
as well as at higher energies 
at RHIC~\cite{v2_rhic}.

\begin{figure}[h]
\vspace*{-0.2cm}
\begin{minipage}[t]{0.48\textwidth}
\centering 
\includegraphics[width=6cm,clip=]{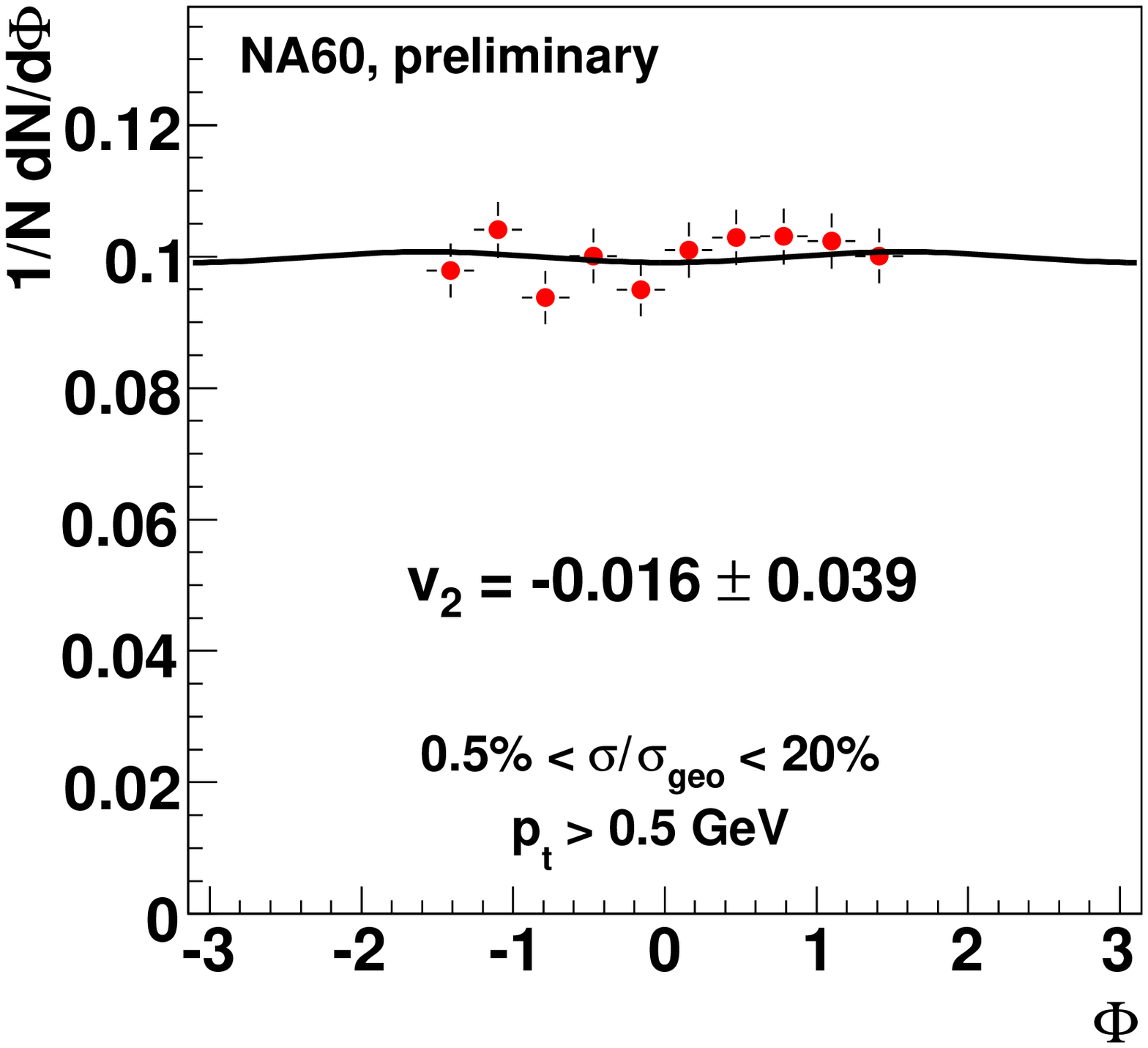}
\end{minipage}
\begin{minipage}[t]{0.48\textwidth}
\centering 
\includegraphics[width=6cm,clip=]{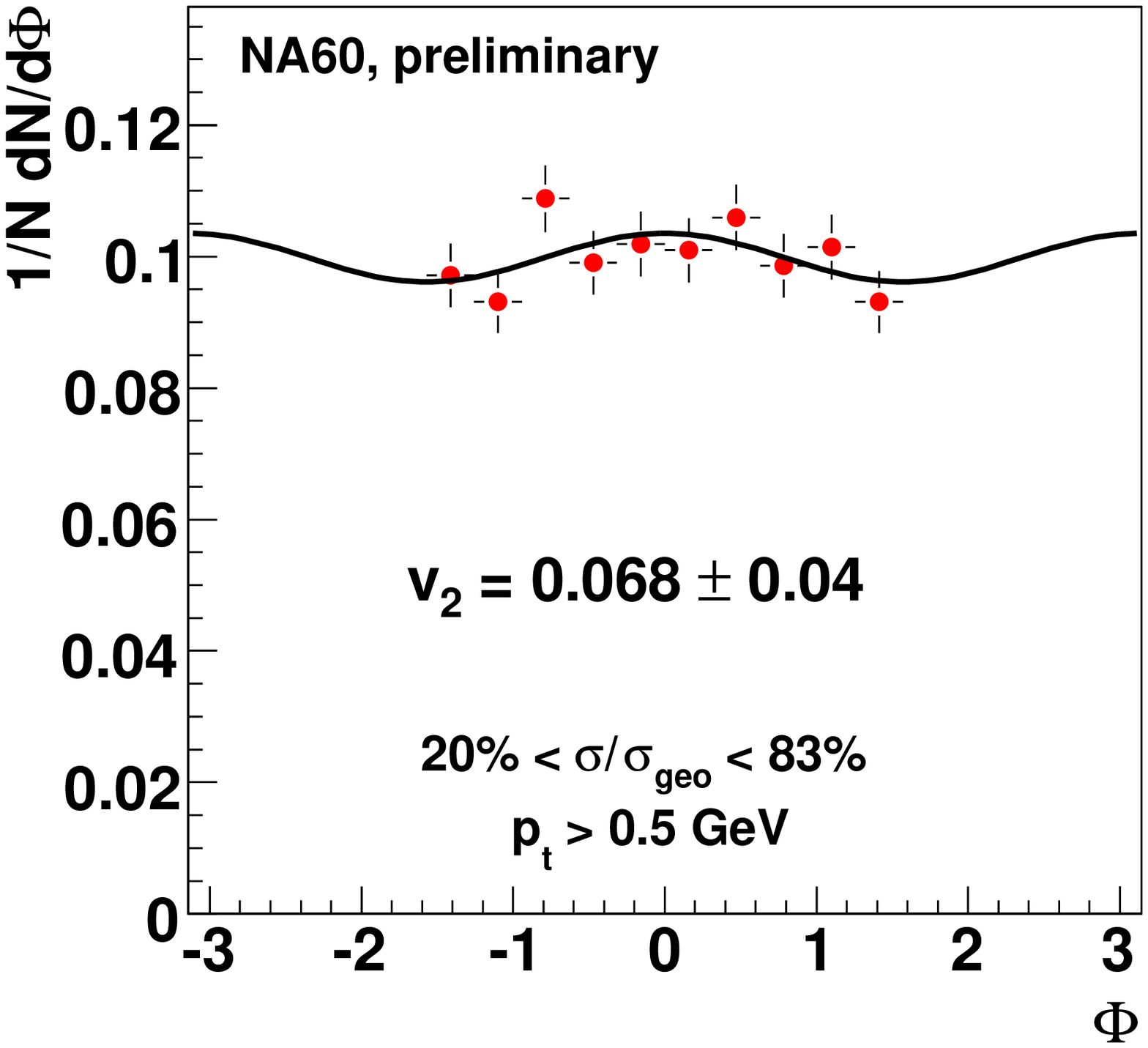}
\end{minipage}
\vspace*{-0.4cm}
\caption{Azimuthal emission patterns of \mbox{J/$\psi$} in \mbox{In+In}
at $158$~$A$~GeV, on the left hand side for more central 
collisions, on the right hand side for more peripheral collisions.} 
\label{fig_phi_fit_JPsi}
\vspace*{-0.2cm}
\end{figure}
Using the energy deposited in the ZDC the data 
can be grouped according 
to the collision centrality. Fig.~\ref{fig_v2_pt_varbin_cent}
shows $v_2$ for charged particles as a function of $p_{\rm t}$ 
for four centrality bins. The quoted fractions 
of the total geometrical cross section have been 
calculated within a Glauber model.
The trend as a function of $p_{\rm t}$  
is similar for all collision centralities with 
the values for $v_2$ increasing for more peripheral
collisions. This can be seen in more detail
in Fig.~\ref{fig_v2_cent_varbin} showing
$v_2$ as a function of the collision centrality
for three bins in $p_{\rm t}$. 

Having determined the orientation of the reaction plane
with the charged particle tracks measured in the
vertex tracker we can determine as well the 
azimuthal angle emission pattern of the \mbox{J/$\psi$}
with respect to this plane.
Figure~\ref{fig_phi_fit_JPsi} shows the azimuthal angle distributions
for \mbox{J/$\psi$}, on the left hand side for the more 
central events (0.5\% $ < \sigma/\sigma_{\rm geo} < $ 20\%),
on the right hand side for the more peripheral events
(20\% $ < \sigma/\sigma_{\rm geo} < $ 83\%).
In total these distributions contain approximately
12000 \mbox{J/$\psi$} which is about 50\%
of the full statistics available for this type of analysis. 
The more peripheral data seem to indicate a non-isotropic 
emission pattern but the limited statistics does not allow
for a solid conclusion up to now.
The analysis of the remaining statistics is under way.


\section*{References}
\begin{thebibliography}{99}
\bibitem{na50setup}L.~Anderson \etal (NA10 Collaboration),
Nucl. Instr. and Meth.~{\bf 223} (1984) 26.
\bibitem{Gluca:2005}G.~Usai \etal (NA60 Collaboration),
Eur. Phys. J.~{\bf C43} (2005) 415.
\bibitem{Keil:2005zq} M.~Keil \etal, 
Nucl. Instrum. Meth.~{\bf A539} (2005) 137  
and {\bf A546} (2005) 448.
\bibitem{Riedler:2006} P.~Riedler \etal, 
Nucl. Instrum. Meth.~{\bf A} in press, available online.
\bibitem{Ruben:2005qm} R.~Shahoyan \etal (NA60 Collaboration), 
Eur. Phys. J.~{\bf C43} (2005) 209.
\bibitem{Aga05} G.~Agakichiev \etal (CERES Collaboration),
Phys. Rev. Lett.~75 (1995) 1272;
Eur. Phys. J.~{\bf C4} 231; 
Eur. Phys. J.~{\bf C41} (2005) 475.
\bibitem{genesis} S.~Damjanovic, A.~De~Falco and H. W\"ohri 
(NA60 Collaboration), NA60 Internal Note 2005-1.
\bibitem{genesis_ceres} 
H.~Sako (CERES Collaboration), GSI Tech. Report 03-24.
\bibitem{Damjanovic:qm2005} E.~Scomparin \etal (NA60 Collaboration),
Proc. Quark Matter 2005, Nucl. Phys.~{\bf A} in print.
\bibitem{prl_lmr} R.~Arnaldi \etal (NA60 Collaboration),
Phys. Rev. Lett.~{\bf 96} (2006) 162302.
\bibitem{Rapp:1995zy} G.~Chanfray, R.~Rapp and J.~Wambach, 
Phys. Rev. Lett.~{\bf 76} (1996) 368;
R.~Rapp, G.~Chanfray and J.~Wambach, Nucl.\ Phys.\ {\bf A617} (1997) 472.
\bibitem{Rapp:1999ej} R.~Rapp and J.~Wambach, 
Adv. Nucl. Phys.~{\bf 25} (2000) 1.
\bibitem{Brown:kk} G.~E.~Brown, M.~Rho, 
Phys. Rev. Lett.~{\bf 66} (1991) 2720 (1991);
G.~Q.~Li, C.~M.~Ko and G.~E.~Brown, Phys. Rev. Lett.~{\bf 75} (1995) 4007.
\bibitem{Brown:2001nh} G.~E.~Brown and M.~Rho,
Phys. Rep.~{\bf 363}, 85 (2002).
\bibitem{Agakichiev:1997au} G.~Agakichiev \etal 
(CERES Collaboration), Phys. Lett.~{\bf B422} (1998) 405; 
B.~Lenkeit \etal, Nucl. Phys.~{\bf A661} (1999) 23c. 
\bibitem{gale:nn} G.~Q.~Li and C.~Gale, 
Phys. Rev. Lett.~{\bf 81} (1998); 
Phys. Rev. C~{\bf 58} (1998) 2914.
\bibitem{na50jpsi} M.~C.~Abreu \etal (NA50 Collaboration),
Phys. Lett.~{\bf B410} (1997) 337;
B.~Alessandro \etal (NA50 Collaboration),
Eur. Phys. J.~{\bf C39} (2005) 335. 
\bibitem{ArnQM} R.~Arnaldi \etal (NA60 Collaboration), 
Proc. Quark Matter 2005, Nucl. Phys.~{\bf A} in print.
\bibitem{Bor05} G.~Borges \etal (NA50 Collaboration), 
Eur. Phys. J.~{\bf C43} (2005) 161.
\bibitem{npart_calc} 
M.~Abreu \etal (NA50 Collaboration),
Phys. Lett. ~{\bf B521} (2001) 195.
\bibitem{poskanzer} A.~M.~Poskanzer, S.~A.~Voloshin, 
Phys. Rev.~{\bf C58} (1998) 1671.
\bibitem{na49} C.~Alt \etal (NA49 Collaboration),  
Phys. Rev.~{\bf C68} (2003) 034903.
\bibitem{ollitrault} J.~Y.~Ollitrault, nucl-ex/97110003.
\bibitem{dinh00}  P.~M.~Dinh, N.~Borghini, J.~Y.~Ollitrault, 
Phys. Lett.~{\bf B477} (2000) 51.
\bibitem{v2_sps} 
J.~Milo{\v s}evi{\'c} \etal (CERES Collaboration), SQM2006, these proceedings.
\bibitem{v2_rhic}
M.~Oldenburg \etal (STAR Collaboration), Quark Matter 2005, 
Nucl. Phys.~{\bf A} in print, nucl-ex/0510026; S.~Adler \etal 
(PHENIX Collaboration), Phys. Rev. Lett.~{\bf 91} (2003) 182301.

\end{thebibliography}
\end{document}